# Fully bio-based Poly (Glycerol-Itaconic acid) as supporter for PEG based form stable phase change materials

Guang-Zhong Yin,[a] José Luis Díaz Palencia,[b] De-Yi Wang [a, *]

[a] IMDEA Materials Institute, C/Eric Kandel, 2, 28906 Getafe, Madrid, Spain
[b] Universidad Francisco de Vitoria, Ctra. Pozuelo-Majadahonda Km 1,800, 28223, Pozuelo de Alarcón, Madrid

Email: deyi.wang@imdea.org

**Abstract**

A novel fully bio-based Poly (Glycerol-Itaconic acid) (PGI) was designed and highly efficiently synthesized by solvent-free polycondensation. The Poly (ethylene glycol) (PEG) was used as the phase change materials (PCM) working substance and encapsulated by the sustainable PGI supporter. PEG chains were tightly encapsulated with the PGI supporting material mainly under hydrogen bonds due to the structural compatibility between PGI and PEG. The PCMs can achieve high form stability and high phase change enthalpies in the same kinds of PCMs. Furthermore, the phase change temperatures and enthalpies of the PCMs can be adjusted conveniently by regulating the PEG content and molecular weight. Notably, this process extremely facilitates the realization of efficient mass production due to the eco-friendly nature, high efficiency and low cost.

## 1. Introduction

Phase change materials (PCMs) were widely used in sustainable thermal energy storage [1], thermal management [2], waste heat recovery [3] and solar energy harvesting. [4] Many studies focused on the exploration of the diversity of PCMs with desirable properties, such as high form stability, high phase change enthalpy and high thermal conductivity. [5, 6] Generally, an encapsulation method has been used to prepare form stable PCMs for resolving the leakage problem. Form stable PCMs usually include a working substance (e.g. paraffin, stearic acid, and lauryl alcohol) and a supporting material. Polyethylene glycol (PEG) was widely selected as the working substance because of its inexpensive nature, non-volatility, high melting enthalpy, [7] high thermal stability, excellent cycle performance, and adjustable phase change temperature due to the wide range of melting temperatures with a different molecular weight.

There are several methods to resolve the problem of poor form stability of the PEG based PCMs, and these methods are mainly related to fabricating nano- or micro-capsules and constructing three-dimensional framework structure. Some examples of form stabilized PCM of utilizing PEG were widely reported, such as PEG/graphene oxide,[8] PEG/expanded graphite, [9, 10] PEG/diatomite, [11] PEG/silica, PEG/silica fume composites, [12] [13] PEG/active carbon [14], and PEG/cellulose [15, 16] or some other bio-based supporter [17] have also been reported to prevent PEG from leakage. Although fabricating nano-/micro-capsules method provides good encapsulation stability, the preparation cost of this kind of PCMs is relatively high. [16] Moreover, traditional methods usually have complex process, non-biomass, and long production cycle. No matter what kind of stabilizing strategy, mass production is the cornerstone of application and promotion. According to the current research situation, it is relatively simple to encapsulate PCMs based on bio-based porous skeleton and organic 3D network structure because only physical mixing/adsorption and impregnation process are usually required. The relatively simple preparation process is conducive to the mass production. However, the synthesis of form stable PCMs by using core–shell encapsulation, chemical crosslinking or copolymerization method is commonly accompanied by a series of chemical or physical process. Therefore, the methods mentioned above are usually found to be unsuitable for commercial upscaling. Hence, it could be further optimized on the design and synthetic methods of encapsulation to improve the process parameter and simplify the process route. [18]

At present, the utility of environment-friendly materials is necessary for securing a sustainable future against climate change and growing energy demand. [19] As outlined in the 2020 issued European Industrial Strategy, the world needs industry to become greener, more circular, and more digital while remaining competitive on the global stage. [20] With respect to the chemicals market, more and more chemicals and materials production in Europe need to be bio-based (25% by 2030 compared to 10% in 2010). [21] Thus, the development of facile, highly efficient, and sustainable composites for PCMs is an essential route for their large-scale production and eco-friendly to the world. [22]

Consequently, it would be of great importance to develop convenient and efficient methods to prepare a form stable bio-based PCM. In this work, we designed and prepared a novel PCM by using PEG as a work substance, which was tightly encapsulated with a fully biomass phase change supporting material, Poly (Glycerol-Itaconic acid) (PGI), under the action of hydrogen bond. PGI can be polymerized by Glycerol and Itaconic acid, which exhibit good compatibility with PEG. Owing to the molecular compatibility between the supporting material PGI and PEG, it is expected that the composites present good form stability and high thermal storage performances.

## 2. Results and discussion

### 2.1 Solvent-free method for synthesis of PGI/PEG PCM composites

PEG 6k was selected as PCM work substance in current study. PGI encapsulated PEG 6k can be prepared by using the fully-biobased PGI via two-step method under solvent-free conditions. The materials preparation and characterization methods are all provided in Supplementary materials. Notably, the method here is solvent free, which is convenient for construction in practical applications. **Figure 1a** showed the step 1 for PGI/PEG precursor synthesis. Typically, Glycerol, Itaconic acid and PEG 6k were mixed together and heat at 180 °C for 6 h, giving rise to a light-yellow viscous oil. There should be a small proportion of PEG that was grafted on PGI framework due to the esterification between the PEG terminals (-OH) and carboxyl from Itaconic acids. This small amount of PGI-PEG can have a positive effect on the stabilization of PEG during melting. We have characterized the chemical structure of raw materials and PGI/PEG precursors by $^{13}$C NMR. Sample PGI-45 was selected as a typical example for structure analysis, and the full was shown in **Figure 1b**. The carbon signals from Itaconic acid are clearly assigned (Figure 1b), which are at ~$\delta$170 ppm (peak a+b), $\delta$ 127.41-135.26 ppm (peak c+d) and $\delta$ 36.94 ppm (peak e), respectively. The assignments of the glycerol carbons and the glyceryl units are also shown in Figure 1b. The observed chemical shift information listed in Supplementary materials (Figure S1-S3) are consistent with those previously reported. [23-25] Through the signal attribution of Figure S3, we know that the glyceryl component of PGI in the prepolymer exists in five forms as shown in Figure 1b inset, namely, G (Glycerol residual), $T_G$, $T_{13}$, $L_{13}$, and $L_{1,2}$. Based on the chemical analysis, we can assign that the Itaconic acid related structure and G, $T_G$, $T_{13}$, $L_{13}$ and $L_{12}$ coexist randomly with proper proportions in the PGI/PEG precursors. **Figure 1c** illustrated the PGI/PEG precursor before curing. Under heating and with the mixture of 0.5 wt.% Benzoyl peroxide (BPO) initiator, the entire system formed an isotropic integrated material after vinyl cross-linking (**Figure 1d**). **Figure 1e** shows the microstructure illustration of PCM after cooling. As PEG melts and crystallizes in confined space, the crystallinity and latent heat of phase transformation are affected by PGI framework to some extent, which will be discussed in section 2.3. The PEG chains were encapsulated in the PGI framework driven by the hydrogen bond due to structural compatibility, which will be confirmed in the following section.

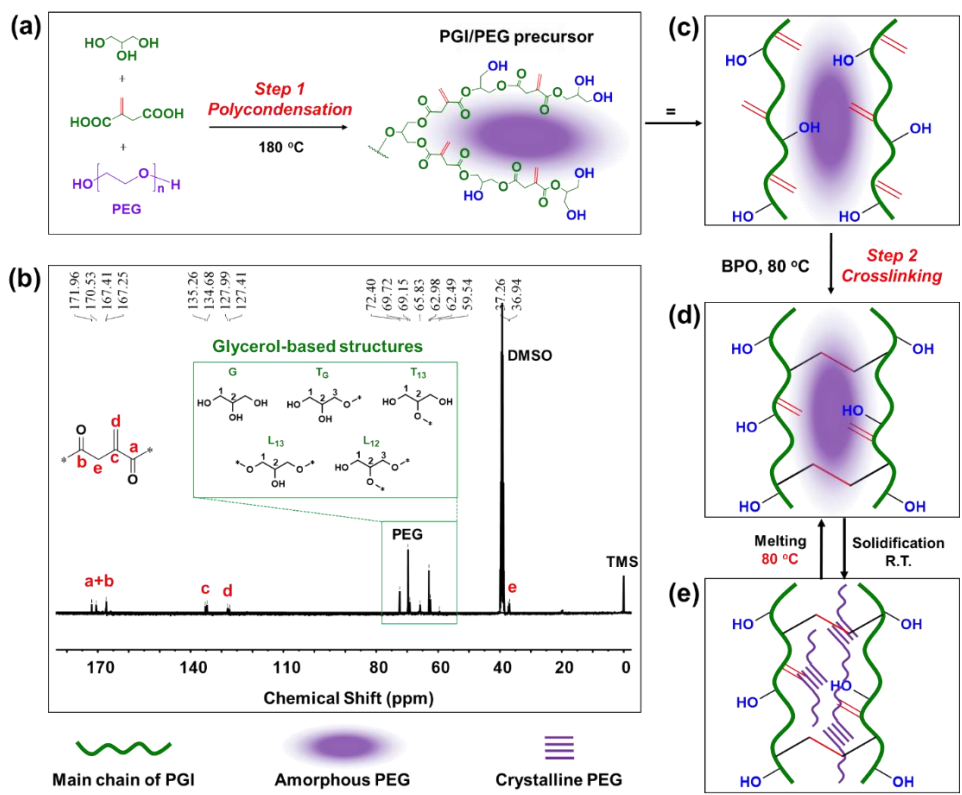

Figure 1. (a) Synthesis of PGI/PEG precursor by polycondensation (Step 1), (b) $^{13}$C NMR full spectrum of PGI/PEG (PGI-45) (Before curing), (c) Schematic diagram of PGI/PEG precursor with free Vinyl and -OH groups, (d) Cured PGI/PEG induced by BPO initiator at 80 °C (Step 2) and (e) structure illustration the reversible phase change.

The FTIR curves of the PGI/PEG composites were carried out and presented in Figure 2a. Pure PEG (Figure S4) shows the typical peaks at ~3440 cm$^{-1}$ for terminal -OH groups and 2876 cm$^{-1}$ for -CH$_2$-.[26] $I_{C=O}$ is the relative intensity of peak at ~1720 cm$^{-1}$ (**Figure 2b**). It is selected as the internal standard of FTIR analysis. The $I_{Vinyl}/I_{C=O}$ is the intensity ratio of vinyl characteristic signal (~1640 cm$^{-1}$) and peak C=O. As it can be seen, the $I_{Vinyl}/I_{C=O}$ for PGI-30 (before curing) and PGI-30 (after curing) are 0.3504 and 0.1798, respectively, which indicated the crosslinking occurred under the action of BPO initiator. The second difference is the shift of hydroxyl peak from 3412 cm$^{-1}$ to 3436 cm$^{-1}$, a higher position (**Figure 2c**). The blue shift occurs as a result of reduction in the percentage of self-hydrogen bonded -OH groups because of the formation of intermolecular hydrogen bonds. [27, 28] These intermolecular interactions will fix the PEG chains to the PGI frameworks. Therefore, PGI/PEG composites undergo solid-solid phase change.

In **Figure 2d**, the peaks of PEG appeared at ~19° (120) and ~23° (032)/(112). [1] The typical peak of PGI/PEG PCMs are quite close to those of PEG, indicating that the encapsulation of PEG and PGI does not influence their respective crystal of PEG significantly.

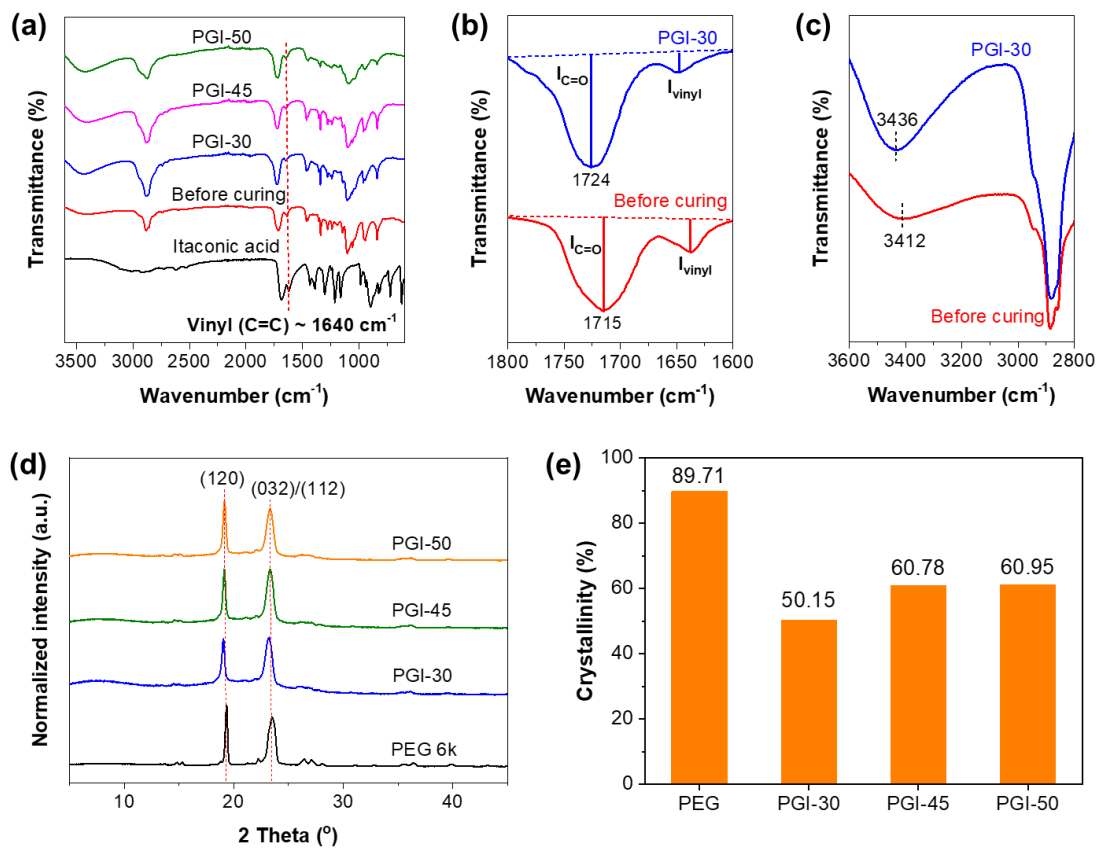

Figure 2. (a) FTIR spectra of Itaconic acid, PGI/PEG precursor and the cured samples, (b) selected region of FT-IR spectra of PGI-30 before and after curing to show the relative intensity of Peak C=O and Peak Vinyl, $I_{vinyl}/I_{C=O}$ (before curing)=0.3504, $I_{vinyl}/I_{C=O}$ (PGI-30)= 0.1798, (c) wavenumber shift of hydroxyl group peak, (d) XRD patterns of PEG 6k, PGI-30, PGI-45 and PGI-50, and (e) the crystallinity (%) results of the PEG 6k, PGI-30, PGI-45 and PGI-50 obtained according to DSC results (second heating, 10 ºC min$^{-1}$).

According to DSC results, we can calculate the crystallinity % via equation (S1, Supporting information). As shown in **Figure 2e**, pure PEG 6k has the highest crystallinity with value of 89.71%. After the encapsulation, crystallinity decreased significantly. Typically, the crystallinity of PGI-30, PGI-45 and PGI-50 are 50.15%, 60.78% and 60.95%, respectively. The decrease of crystallinity may be due to the inhibition effect of PGI framework as impurities on the crystallization of PEG. Specifically, the hydrogen bonds between PGI and PEG as well as glycerol and PEG and free glycerol play an inhibitory role in the complete crystallization of PEG.

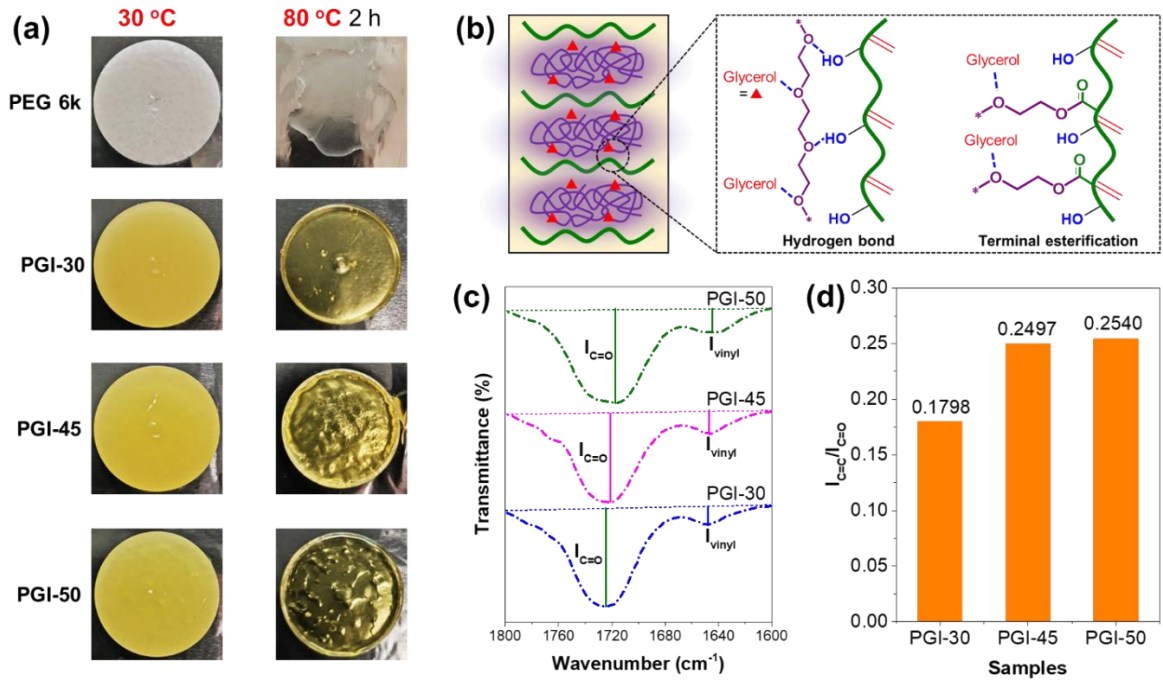

Figure 3. (a) Form stability test: images of PGI/PEG PCMs with different PEG loads at room temperature and 80 °C for 2 hours, (b) schematic illustration showing the compatibility based on the hydrogen bond or chemical grafting via terminal of PEG esterification with the -COOH in PGI, (c) FTIR curves for calculating the relative intensity of $I_{vinyl}/I_{C=O}$, and (d) the trend of $I_{vinyl}/I_{C=O}$ values.

*2.2 Form stability*

The PGI/PEG PCMs with different PEG 6k contents (62.76 %–72.67%) were prepared and tested for their form stabilities. PEG 6k and PGI/PEG PCMs were heated at 80 °C for 2 h. As shown in **Figure 3a**, PEG 6k completely melted, while the PGI/PEG PCMs were only slightly softened, remained in a stable and solid state. However, the leakage occurred when the load content reached 75.24 % (namely, sample PGI-55, Figure S5). Thus, the load of PEG should be lower than 75.24 %.

**Figure 3b** schematically illustrated the compatibility based on the hydrogen bond between PEG and PGI, and esterification between terminal hydroxyl of PEG and the carboxyl in PGI. There are abundant of hydroxyl groups on the PGI chains, which may interact with PEG through a hydrogen bonding interaction (as discussed in Section 2.1), giving rise to the improvement of the encapsulation of PEG during the phase transition, which is in good accordance with the previous reports. [29] Moreover, due to the corresponding steric hindrance introduced by PEG, the density of double bond is reduced, so that the crosslinking degree decreased with the increase of PEG contents under the same crosslinking conditions. Specifically, as shown in Figure 3c and 3d, the specific intensity ratio of $I_{vinyl}/I_{C=O}$ increased with the increasing of PEG content, which indicates that the proportion of residual vinyl group is increasing. Consequently, the higher contents of vinyl groups remained, directly giving rise to the decrease of crosslinking degree, which is also the direct reason of leakage during phase transition for samples PGI-55.

Table 1. Some parameters of the PCMs.

| Samples | PEG contents (%) | $\Delta H_m$ (J g$^{-1}$) | $\Delta H_s$ (J g$^{-1}$) | $\Delta E_a$ (kJ mol$^{-1}$) | Extent of Supercooling (°C) | Heat loss (%) | Form stability (Leakage) |
|---|---|---|---|---|---|---|---|
| PEG 6k | - | 176.20 | - | 561 | 17.9 | 3.35 | Yes |
| PGI-PEG 30 | 62.76 | 61.81 | 59.03 | 199 | 13.3 | 4.50 | No |
| PGI-PEG 45 | 68.70 | 82.01 | 78.35 | 166 | 13.1 | 4.46 | No |

| PGI-PEG 50 | 72.67 | 86.93 | 83.65 | 205 | 13.2 | 3.77 | No |

Note: the extent of supercooling and heat loss were both calculated according to the first cycle with heating rate of 10 °C min$^{-1}$.

## 2.3 Thermal response and cycle performance

The $\Delta E_a$ of nonisothermal phase transition can be calculated by Kissinger's Equation: [30]

$$\frac{d[ln(\varphi/T_P^2)]}{d(\frac{1}{T_P})} = -\frac{\Delta E_a(T)}{R} \tag{1}$$

where $\varphi$ is the heating rate, $T_P$ is the peak temperature, and R is the gas constant. As a consequence, $\Delta E_a$ can be calculated through the slope of the curve of $ln\left(\frac{\varphi}{T_P^2}\right)$ against $\frac{1}{T_P}$. **Figure 4a** shows the nonisothermal melting peaks of PGI-30 determined at different heating rates. The DSC curves of other samples (PEG 6k, PGI-45 and PGI-50) are shown in Figure S5, S6 and S7, respectively. **Figure 4b** presents the plots based on the data corresponding to **Figure 4a** and Figure S5-7. The following activation energy values were obtained: $E_{a, PEG}$=561 kJ mol$^{-1}$ for neat PEG, $E_{a, PGI-30}$=199 kJ mol$^{-1}$, $E_{a, PGI-45}$=166 kJ mol$^{-1}$, $E_{a, PGI-50}$=205 kJ mol$^{-1}$ for PGI-30, PGI-45 and PGI-50, respectively. The activation energies for all the samples are also listed in **Table 1** and shown in **Figure 4c**. It's found that all the $\Delta E_a$ of the PGI/PEG PCMs are much lower than that of pure PEG, which is mainly because the -OH in free glycerol (Figure S3 showed that some free Glycerol remained in the system) might increase the mobility process of PEG chain in PCM to a certain extent.

The extent of supercooling (ΔT, °C) was evaluated by the following equation [31, 32] and listed in Table 1:

$$\Delta T = T_{m,onset} - T_{s,onset} \qquad (2)$$

The decrease of the extent of supercooling indicates that the PGI can promote the nucleation process of PEG molecular chain. It is reported that the phase transition process of pure PEG is dominated by homogenous nucleation and growth mechanism. While, PGI or Glycerol can provide numerous heterogeneous nucleation sites for the phase transition of PEG. Typically, PGI or Glycerol can serve as intramolecular heterogeneous nucleation agents during the phase transformation process of PCMs.

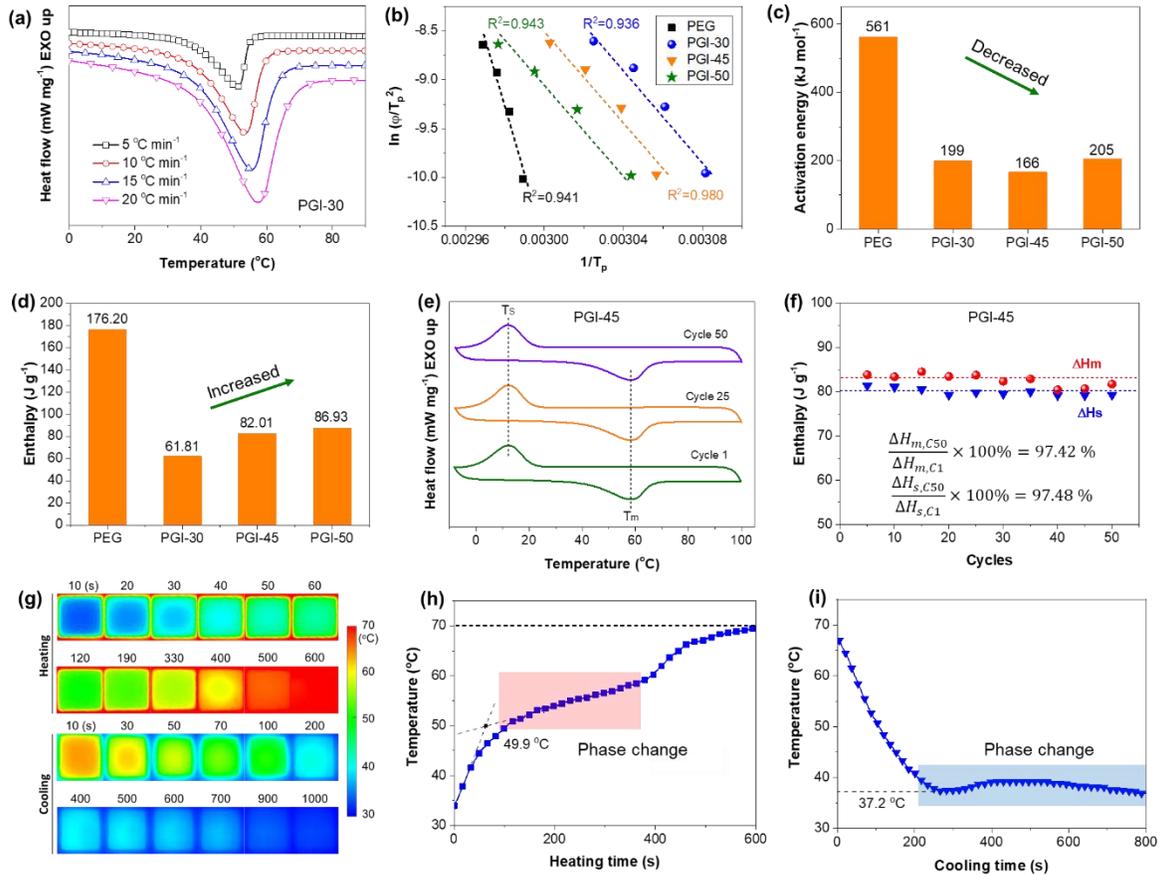

**Figure 4**. (a) Nonisothermal DSC curves of sample PGI-30 with different heating rates, (b) $\ln(\varphi/T_p^2)$ versus $1/T_p$ plots, (c) $\Delta E_a$ of PEG 6k, PGI-30, PGI-45 and PGI-50, (d) melting

enthalpies of PEG 6k and PCMs with different PEG 6k contents (10 ºC min$^{-1}$), (e) DSC cycle curves of PCMs, (f) the cycle performance (including the melting enthalpy, $\Delta H_m$, and solidification enthalpy, $\Delta H_s$) of PGI-45 as a typical example, (g) thermal images of PGI-50 during heating and cooling, (h) the heating curves and (i) cooling curves of the PGI-50 samples.

Furthermore, the percentage of heat loss ($\eta$, %) was evaluated by equation 3: [33]

$$\eta = \frac{\Delta H_m - \Delta H_s}{\Delta H_m} \times 100 \% \tag{3}$$

where, $\Delta H_m$ was the melting enthalpy of the PCMs, and $\Delta H_s$ is solidification enthalpy. As shown in **Table 1** and **Figure 4d**, the PCMs exhibited a phase change enthalpy (61.81~86.93 J g$^{-1}$). Furthermore, according to Equation (3), the heat loss for pure PEG and PCMs are calculated (**Table 1**). All the calculated heat loss are quite low (<4.50 %). The PCMs had high cycle performance because the main parameters for PCM, such as $\Delta H_m$, $\Delta H_s$ and melting temperature, et al., remained virtually unchanged (**Figure 4e, f,** and **Table S2**). The PGI/PEG composites are therefore regarded as highly reversible, form stable PCMs for thermal energy storage. We performed heat response test on sample PGI-50. The data are shown in **Figure 4 g**, **h**, and **i**. Under the experimental test conditions, (the specific size and test method are indicated in Supplementary materials) the sample started the phase transition at ~ 49.9 ºC, and the temperature is maintained at ~55 ºC for about 400 s. During the cooling process, the phase transition begins at ~ 37.2 ºC and the temperature is maintained at 35 ~ 40 ºC for more than 10 minutes. It is worth noting that the extent of supercooling of the sample can be estimated to be about 12.7 ºC, which is in good agreement with the results obtained by DSC data (Table 1).

Solvent free two-steps (polycondensation and casting-curing) method with high efficiency were used to prepare the PGI/PEG PCMs. The precursors have good fluidity at 80 ºC, are suitable for casting in various cavities, and can be cured at 80 ºC overnight. Therefore, it can be assigned to be a convenient and efficient pathway for construction during practical application. On the basis of easily adjusting the enthalpy of phase transition, we can also control the phase transition temperature by changing the molecular weight of PEG. PEG 20 kg mol$^{-1}$ was selected as a typical example to confirm this statement. As shown in Figure S9, the phase transition temperature increases from 55.9 ºC of PGI-45-PEG 6k to 59.3 ºC of PGI-45-PEG 20k. The cooling phase transition increases from $T_{s,\,onset}$ = 27.35 ºC and $T_{s,\,peak}$ = 13.04 ºC of PGI-45-PEG 6k to $T_{s,\,onset}$ = 36.15 ºC and $T_{s,\,peak}$ = 30.78 ºC of PGI-45-PEG 20k, respectively. The melting enthalpies of phase change are close. Concretely, PGI-45 PEG 6k is 82.01 J g$^{-1}$, and PGI-PEG 20k is 81.93 J g$^{-1}$. From the comparison, we found that the higher the molecular weight of PEG, the higher the phase transition temperature, which is consistent with the results reported in the literature elsewhere. [7] The $T_{m,\,onset}$ of the composite PCMs was 37.86~41.92 ºC, and the corresponding melting enthalpy was 61.80-86.93 J g$^{-1}$. In addition, it was very stable after 50 times heating/cooling cycles during long-term application. In a word, the PGI/PEG composite PCM had good performance and was suitable for application in building energy conservation.

Some methods, solvents, and enthalpies of phase transition of bio-based support for PEG PCMs are listed in Table S4. We also listed some other methods for preparation of form stable PCMs (Table S5). The traditional methods mainly contain the solvent (water, acetone, DMF, Toluene, etc.), which greatly increases the preparation period, and it is difficult to achieve controllable pouring molding in the practical application process such as wall insulation. Furthermore, it is worth noting that the phase transition enthalpy of this study is at a high level in the same kind of PCM. As well known, nowadays the design of eco-friendly, value-added, conveniently available, and cost-effective materials is crucial in the large-scale and efficient application of energy storage and conversion. Fortunately, the method here is totally solvent-free and can be conveniently casted and molded. As shown in Table S6, the raw materials for PGI have quite low cost and are fully biomass. Therefore, it is expected the PCMs in this work will have its practical application in the future.

## 3. Conclusion

Novel sustainable phase change composite materials were designed and prepared by using PEG as a work substance, which was tightly encapsulated with crosslinked PGI under hydrogen bond. The fully biobased PGI can support up to 72.67 % of PEG 6k. The PGI/PEG composite PCMs have excellent form stabilities and relative high phase transition enthalpies (61.8-86.9 J g$^{-1}$) in the same kinds of PCMs. Through structural analysis, the main reason for the form stability is mainly intermolecular hydrogen bonds, giving rise to the leakage proof performance. Furthermore, the phase change temperatures and enthalpies of the PCMs can be adjusted conveniently by regulating the PEG content and molecular weight. It is worth emphasizing that the process facilitates the realization of efficient mass production due to its sustainable nature (being solvent free and eco-friendly), high efficiency, and low cost.

## References

[1] Z. Shi, H. Xu, Q. Yang, C. Xiong, M. Zhao, K. Kobayashi, T. Saito, A. Isogai, Carboxylated nanocellulose/poly(ethylene oxide) composite films as solid–solid phase-change materials for thermal energy storage, Carbohyd. Polym. 225 (2019) 115215.

[2] B. Wang, G. Li, L. Xu, J. Liao, X. Zhang, Nanoporous Boron Nitride Aerogel Film and Its Smart Composite with Phase Change Materials, ACS Nano 14(12) (2020) 16590-16599.

[3] J. Huang, B. Wu, S. Lyu, T. Li, H. Han, D. Li, J.-K. Wang, J. Zhang, X. Lu, D. Sun, Improving the thermal energy storage capability of diatom-based biomass/polyethylene glycol composites phase change materials by artificial culture methods, Sol. Energ. Mat. Sol. C. 219

(2021) 110797.

[4] D. Liu, C. Lei, K. Wu, Q. Fu, A Multidirectionally Thermoconductive Phase Change Material Enables High and Durable Electricity via Real-Environment Solar–Thermal–Electric Conversion, ACS Nano 14(11) (2020) 15738-15747.

[5] Y. Zhang, K. Sun, Y. Kou, S. Wang, Q. Shi, A facile strategy of constructing composite form-stable phase change materials with superior high thermal conductivity using silicagel industrial wastes, Sol. Energy 207 (2020) 51-58.

[6] A. Bashiri Rezaie, M. Montazer, Shape-stable thermo-responsive nano Fe3O4/fatty acids/PET composite phase-change material for thermal energy management and saving applications, Appl. Energ. 262 (2020) 114501.

[7] Y. Meng, Y. Zhao, Y. Zhang, B. Tang, Induced dipole force driven PEG/PPEGMA form-stable phase change energy storage materials with high latent heat, Chem. Eng. J. 390 (2020) 124618.

[8] G.-Q. Qi, C.-L. Liang, R.-Y. Bao, Z.-Y. Liu, W. Yang, B.-H. Xie, M.-B. Yang, Polyethylene glycol based shape-stabilized phase change material for thermal energy storage with ultra-low content of graphene oxide, Sol. Energ. Mat. Sol. C. 123 (2014) 171-177.

[9] Y. Yang, Y. Pang, Y. Liu, H. Guo, Preparation and thermal properties of polyethylene glycol/expanded graphite as novel form-stable phase change material for indoor energy saving, Mater. Lett. 216 (2018) 220-223.

[10] X. Chen, H. Gao, M. Yang, W. Dong, X. Huang, A. Li, C. Dong, G. Wang, Highly graphitized 3D network carbon for shape-stabilized composite PCMs with superior thermal energy harvesting, Nano Energy 49 (2018) 86-94.

[11] T. Qian, S. Zhu, H. Wang, B. Fan, Comparative Study of Carbon Nanoparticles and Single-Walled Carbon Nanotube for Light-Heat Conversion and Thermal Conductivity Enhancement of the Multifunctional PEG/Diatomite Composite Phase Change Material, ACS Appl. Mater. Interfaces 11(33) (2019) 29698-29707.

[12] C.L. Wang, K.L. Yeh, C.W. Chen, Y. Lee, H.L. Lee, T. Lee, A quick-fix design of phase change material by particle blending and spherical agglomeration, Appl. Energ. 191 (2017) 239-250.

[13] X. Shi, M.R. Yazdani, R. Ajdary, O.J. Rojas, Leakage-proof microencapsulation of phase change materials by emulsification with acetylated cellulose nanofibrils, Carbohyd. Polym. 254 (2021) 117279.

[14] L. Feng, J. Zheng, H. Yang, Y. Guo, W. Li, X. Li, Preparation and characterization of polyethylene glycol/active carbon composites as shape-stabilized phase change materials, Sol. Energ. Mat. Sol. C. 95(2) (2011) 644-650.

[15] C. Lei, K. Wu, L. Wu, W. Liu, R. Du, F. Chen, Q. Fu, Phase change material with anisotropically high thermal conductivity and excellent shape stability due to its robust cellulose/BNNSs skeleton, J. Mater. Chem. A 7(33) (2019) 19364-19373.

[16] X. Wei, X.-z. Jin, N. Zhang, X.-d. Qi, J.-h. Yang, Z.-w. Zhou, Y. Wang, Constructing cellulose nanocrystal/graphene nanoplatelet networks in phase change materials toward intelligent thermal management, Carbohyd. Polym. 253 (2021) 117290.

[17] H. Fang, J. Lin, L. Zhang, A. Chen, F. Wu, L. Geng, X. Peng, Fibrous form-stable phase change materials with high thermal conductivity fabricated by interfacial polyelectrolyte complex spinning, Carbohyd. Polym. 249 (2020) 116836.

[18] K. Yu, Y. Liu, Y. Yang, Review on form-stable inorganic hydrated salt phase change materials: Preparation, characterization and effect on the thermophysical properties, Appl. Energ. 292 (2021) 116845.

[19] M. Wang, J. Pan, M. Wang, T. Sun, J. Ju, Y. Tang, J. Wang, W. Mao, Y. Wang, J. Zhu, High-Performance Triboelectric Nanogenerators Based on a Mechanoradical Mechanism, ACS Sustain. Chem. Eng. 8(9) (2020) 3865-3871.

[20] A New Industrial Strategy for Europe, https://eur-lex.europa.eu/legal-content/EN/TXT/PDF/?uri=CELEX:52020DC0102&from=EN.

[21] BIO-BASED INDUSTRIES for Development & Growth in Europe, Strategic Innovation and Research Agenda (SIRA) (May 2017), https://biconsortium.eu/sites/biconsortium.eu/files/downloads/SIRA-2017-Web.pdf.

[22] D.G. Atinafu, S. Jin Chang, K.-H. Kim, S. Kim, Tuning surface functionality of standard biochars and the resulting uplift capacity of loading/energy storage for organic phase change materials, Chem. Eng. J. 394 (2020) 125049.

[23] M. Agach, S. Delbaere, S. Marinkovic, B. Estrine, V. Nardello-Rataj, Synthesis, characterization, biodegradability and surfactant properties of bio-sourced lauroyl poly(glycerol-succinate) oligoesters, Colloid. Surface. A. 419 (2013) 263-273.

[24] J.-F. Stumbé, B. Bruchmann, Hyperbranched Polyesters Based on Adipic Acid and Glycerol, Macromol. Rapid Comm. 25(9) (2004) 921-924.

[25] T. Zhang, B.A. Howell, A. Dumitrascu, S.J. Martin, P.B. Smith, Synthesis and characterization of glycerol-adipic acid hyperbranched polyesters, Polymer 55(20) (2014) 5065-5072.

[26] S. Karaman, A. Karaipekli, A. Sarı, A. Biçer, Polyethylene glycol (PEG)/diatomite composite as a novel form-stable phase change material for thermal energy storage, Sol. Energ. Mat. Sol. C. 95(7) (2011) 1647-1653.

[27] P. Liu, W. Chen, C. Liu, M. Tian, P. Liu, A novel poly (vinyl alcohol)/poly (ethylene glycol) scaffold for tissue engineering with a unique bimodal open-celled structure fabricated using supercritical fluid foaming, Sci. Rep. 9(1) (2019) 9534.

[28] G. Chen, N. Chen, Q. Wang, Preparation of poly (vinyl alcohol)/ionic liquid composites with improved processability and electrical conductivity for fused deposition modeling, Mater. Design 157 (2018) 273-283.

[29] X. Du, M. Zhou, S. Deng, Z. Du, X. Cheng, H. Wang, Poly(ethylene glycol)-grafted nanofibrillated cellulose/graphene hybrid aerogels supported phase change composites with superior energy storage capacity and solar-thermal conversion efficiency, Cellulose 27(8) (2020) 4679-4690.

[30] X. Chen, H. Gao, G. Hai, D. Jia, L. Xing, S. Chen, P. Cheng, M. Han, W. Dong, G. Wang, Carbon nanotube bundles assembled flexible hierarchical framework based phase change material composites for thermal energy harvesting and thermotherapy, Energy Storage Mater. 26 (2020) 129-137.

[31] Q. Sun, H. Zhang, J. Xue, X. Yu, Y. Yuan, X. Cao, Flexible phase change materials for thermal storage and temperature control, Chemical Engineering Journal 353 (2018) 920-929.

[32] G.-Z. Yin, J. Hobson, Y. Duan, D.-Y. Wang, Polyrotaxane: New generation of sustainable, ultra-flexible, form-stable and smart phase change materials, Energy Storage Mater. 40 (2021) 347-357.

[33] X. Zhang, H. Liu, Z. Huang, Z. Yin, R. Wen, X. Min, Y. Huang, Y. Liu, M. Fang, X. Wu, Preparation and characterization of the properties of polyethylene glycol @ $Si_3N_4$ nanowires as phase-change materials, Chem. Eng. J. 301 (2016) 229-237.